\documentclass[titlepage]{article}
\usepackage{fullpage}
\usepackage{amsmath}
\usepackage{orcidlink}
\usepackage{hyperref}

\DeclareMathOperator{\sgn}{sgn}

\begin{document}

\title{Loss Functions for Measuring the Accuracy of Nonnegative Cross-Sectional Predictions}
\author{Charles D. Coleman 
\orcidlink{0000-0001-6940-8117}\thanks{ This paper reports the general results of research originally undertaken while the author was employed by the Census Bureau.  The views expressed are attributable to the author and do not necessarily reflect those of the Census Bureau.  I would like to thank Jay Siegel, Stan Smith, Bashiruddin Ahmed, Gregg Diffendal and Dave Word for comments and Reuben A. Coleman for research assistance. An earlier version of this paper, entitled “Metrics for Assessing the Accuracy of Cross-Sectional Estimates and Forecasts,” was presented to the Southern Demographic Association meetings in San Antonio, TX, October 1999.}\\
Timely Analytics, LLC \\
E-mail: info@timely-analytics.com}

\maketitle

\abstract{
Measuring the accuracy of cross-sectional predictions is a subjective problem.  Generally, this problem is avoided.  In contrast, this paper confronts subjectivity up front by eliciting an impartial decision-maker’s preferences.  These preferences are embedded into an axiomatically-derived loss function, one of the simplest version of which is described.  The parameters of the loss function can be estimated by linear regression.  Specification tests for this function are described.  This framework is extended to weighted averages of estimates to find the optimal weightings.  A special case occurs when the predictions represent resource allocations: the apportionment literature is used to construct the Webster-Saint Lagüe Rule, a particular parametrization of the loss function.  These loss functions are compared to those existing in the literature. Finally, a family of bias measures are created using signed versions of these loss functions.}

\section{Introduction}

The accuracy of cross-sectional predictions is of importance to a large number of their users: governments, investors, and so on. These data include, for example, population, employment or income per capita by geographic area.  The outstanding characteristics of cross-sectional data are a great range of values and systematically varying error variances and coefficients of variation. Predictions encompass both forecasts and estimates, the only difference, generally irrelevant to this article, is the date being predicted relative to the date the predictions are made.    In the context of predictions, accuracy is generally a subjective concept.\footnote{Compare Lindley (1953, p. 46): “...I feel that loss is not easily separated in our minds from beliefs.”  Later we develop loss functions to reflect beliefs.}  Even if objective measures of scale can be used, they may not necessarily coincide with the user’s concept of loss.  We axiomatically develop loss functionns,  which measure the “badness” of errors as a function of the sizes of the error and the actual values, or, equivalently, the predicted and actual values.  This approach is in contrast to the common use of measures based on time series.\footnote{Armstrong and Collopy (1992) and Fildes (1992), to give two examples, are quite clear about the time series basis of the error measures they use.}  These measures were developed to measure variations in the level of a single variable over time and are, thus, inapplicable to cross-sectional data.\footnote{NRC (1980), is one of the few examples to use cross-sectional criteria: “‘To minimize errors in allocations’ is a vague statement which allows several interpretations (NRC, 1980, p.86).”  The allocations are revenues directed to states and localities in proportion to their populations. I am indebted to Bashiruddin Ahmed for providing this example.  Spencer (1980, 1985, 1986) also uses cross-sectional criteria.}  Initially, an impartial decision-maker is assumed to exist and asked to specify values of his loss function (disguised) for various combinations of errors and actual values.  The decision-maker is assumed to be interested solely in the overall accuracy of the predictions and not to have an interest in the accuracy for a particular area or set of areas.  The context of the predictions is also assumed fixed.\footnote{I’d like to thank Stan Smith for pointing this out.}  Linear regression is then used to estimate the parameters of the loss function.\footnote{The assumption of an impartial decision-maker and the statistical construction of the loss function provide an answer Spencer’s (1986) criticisms about whose preferences are to be incorporated into a von Neumann-Morgenstern loss function and how.  These constructions are very much along the lines of Wald (1950) and Lindley (1953, 1985).}  An upshot of this process is that, in general, no single “ideal” measure of accuracy exists, as the evaluation of accuracy depends on the evaluator.\footnote{Tukey (1986, p. 408) writes in the context of cross-sectional population estimates: “‘Optimum’ estimates, for example, are \underline{only} optimum under narrow specifications that do not hold exactly in practice.”  (Original underline)}  In the special case in which predictions represent resource allocations, the apportionment literature is used to establish the Webster-Saint Lagüe Rule as the appropriate loss function.

	The use of loss functions to measure the accuracy of cross-sectional predictions is hardly a new idea.  Two commonly used measures, the mean absolute percentage error (MAPE) and mean absolute error use the absolute percentage error and absolute error as loss functions, respectively.  Armstrong (1985, ch. 15) discusses various accuracy measures and the “cost” (i.e., loss) functions underlying them.  None of these measures has an axiomatic basis.  Stanford Research Institute (1974) and Spencer (1980) used utility criteria to create the first axiomatically-based cross-sectional loss functions in the context of allocating general revenue sharing funds.  NRC (1980, p. 87) developed similar loss functions to ours by generalizing from special cases.  Spencer (1985) developed loss function representations of four rules that have been used to apportion the U.S. House of Representatives, one of which is the Method of Major Fractions, also known as the Webster-Saint Lagüe Rule.  Loss functions have been used by the U.S. Census Bureau to compare unadjusted and adjusted census counts to artificial target populations, beginning with the 1990 census.  Mulry and Spencer (1993) is an example of this.

	Section 2 axiomatically constructs the simple loss function that we  use.  It further specifies assumptions and develops one of the simplest loss functions that satisfy those assumptions and has the property that the loss associated with a given absolute relative error\footnote{Equivalently, one can also refer to the absolute percentage error (APE), but this will complicate the arguments made later on.  APE will appear, however, in the example in Section 6.} increases in the actual value.  Section \ref{LFest} discusses estimation of this loss function, along with specification tests.   Different sets of predictions may have systematic, offsetting biases.  The decision-maker may wish to form weighted averages of these predictions to produce a set of new predictions.  Subection \ref{optimal} provides a method to search for the optimal weightings which minimize the total loss function. Section \ref{wsl} argues for the use of the Webster-Saint Lagüe Rule when predictions can be viewed as apportionments.  Given a set of predicted and actual values, the total loss function is the sum of the loss functions for the individual observations. Section \ref{example} provides an example of the use of loss functions, comparing the Webster-Saint Lagüe Rule to the commonly used mean absolute percentage error (MAPE).

	Section \ref{compare} compares the metrics developed in this paper to existing metrics in the literature.  Section \ref{bias} constructs a family of bias measures based on the loss functions. Section \ref{conclude} concludes this paper.

\section{The Loss Function}

This section describes the assumptions used to generate the loss function, $L(P;A)$, where $P$ is the predicted value and $A$ is the actual value.\footnote{Muhsam (1966) was apparently the first to use this form of loss in the context of predicting population.  He was concerned with the loss associated with a point forecast of the population of a single geographic area.  He assumed continuity of loss and a version of Assumption 2 below. I'd like to thank Jay Siegel for this point.}  After the assumptions are made, onr of the simplest forms of $L$ which satisfies them is specified.  Restrictions on the values of the parameters of $L$ which make it increase in $A$ for a given relative error are then specified.  The total loss $\mathcal{L}$ is the sum of the losses associated with each observation $i$: $\mathcal{L} = \sum_i^n L(P_i,A_i)$, where $n$ is the number of observations.\footnote{In terms of Spencer (1986), $L$ is a component loss function and $\mathcal{L}$ is the loss function. The U.S. Census Bureau also uses this terminology for evaluating census adjustments.}  While other forms of $\mathcal{L}$ are possible, the summation corresponds to the concept of additively separable utility and obeys the von Neumann-Morgenstern expected utility axioms.\footnote{That is, loss is the negative of utility.  Additive losses are equivalent to additive utility.  The von Neumann-Morgenstern expected utility axioms imply that the expected utility of a gamble is equal to the utilities of the outcomes multiplied by their probabilities.  In terms of loss, the expected loss when losses of values $L_1$ and $L_2$ occur with probabilities $p_1$ and $p_2$, respectively, is $p_1L_1 + p_2L_2$.  This approach is similar to Lindley’s (1985, p. 59) commandment to “Go forth and maximise your expected utility.”  (Original quotes)  Lindley (1985) builds his book around the concept of making decisions that maximize a decision-maker’s expected utility.  Since expectation requires specifying probability distributions, which we do not do, our commandment can be stated as “Go forth and maximize utility.”\label{utility}}

	The first assumption we make is that $L$ is symmetric in the errors:\newline
\textbf{Assumption 1:} $L(A + \epsilon; A) = L(A - \epsilon; A)$ for all $A > 0$.\newline
This assumption is not as innocuous as it looks.\footnote{The measures of bias mentioned in footnote 1 violate this assumption by design.}  It is quite possible that, at least for some range of $A$, that the decision-maker is not indifferent between positive and negative errors. Consider, for example, William Tell and the apple.  If Tell places the arrow one inch above the apple, he suffers one sort of loss.  If he places the arrow the same distance below the apple, he suffers a much more severe loss.\footnote{I would like to thank Dave Word for providing me with this example.}  However, by not making Assumption 1, the resulting asymmetry complicates the definition of $L$.  The symmetry of $L$ allows us to use the equivalent notation $\ell (A,\epsilon) \equiv L(A + \epsilon; A)$ where $\epsilon\ge 0$.

	The next assumption makes $L$, or, equivalently, $\ell$, increasing in the error $\epsilon$:\newline
\textbf{Assumption 2:}   $\partial\ell/\partial\epsilon > 0$ for all $\epsilon > 0$.\newline
Note that this assumption is stated in terms of $\ell$, rather than $L$.  This assumption is quite intuitive, as it states that more accurate predictions are preferred to less accurate ones.\footnote{It is easy to show that Assumption 2 implies Fisher-consistency (Spencer, 1986, p. 395): $L$ is uniquely minimized when $P_i = A_i$ for all $i$.}

	Finally, we want $L$, or, equivalently, $\ell$, to decrease in $A$.  This means that for a given value of $\epsilon$, the loss associated with it decreases with the actual value.  This has two justifications.  First, for example, an error of 500 when the true value is 1,000 is a whopping 50\%, a highly significant error.  However, the same error, when the true value is 1,000,000 is akin to a roundoff error.   In short, error variance increases in the actual value.  Second, when making predictions, the coefficient of variation of the errors, $\sigma/\mu$, where $\sigma$ is the variance and $\mu$ is the expected value, generally decreases in $A$.7  We state this formally as:\newline
\textbf{Assumption 3:} $\partial\ell/\partial A > 0$  for all $A > 0$.\newline
	One of the simplest function swhich satisfies Assumptions 1–3 and admits Property 1 below is the Cobb-Douglas function
\begin{subequations}
	\begin{equation}
		\epsilon^p A^q
           \end{equation}
or, equivalently,
	\begin{equation}
		|P-A|^p A^q \label{PAloss}
           \end{equation}
where $\epsilon, p > 0$ and $q < 0.$\footnote{NRC (1980, p. 87) obtains the same loss function by generalizing from special cases.} Note that $\epsilon = |P - A|$.\label{epsilon}
	The loss function (1a) can be interpreted as an exponentially weighted product of the absolute value of the difference and the absolute relative difference:
	\begin{equation}
		|P - A|^{p+q} \left|\frac{P - A}{A}\right|^{-q}.
        \end{equation}
\end{subequations}
In order for the loss function to increase in the absolute value of the difference $|P - A|$, the sum $p + q$ must be positive.  When this is true, the loss function gains the desirable property that it rises in $A$ for a given absolute relative error.  We formalize this argument as follows.  The absolute relative error is: 
\begin{equation}
	\left|\frac{P - A}{A}\right|
\end{equation}
Note that, in this case, $p = 1, q = -1$ and $p + q = 0$.  Choosing $q >-1$ makes the corresponding ($p = 1$) loss function increase in $A$.  Therefore, $p + q > 0$ in the loss function when $p = 1$.  We generalize the argument by raising equation (2) to any positive power $r$:
\begin{equation}
	\left|\frac{P - A}{A}\right|^r
\end{equation}
Here $p = r$ and $q = -r$ and, again, $p + q = 0$.  Choosing $q > -r$ makes the corresponding ($p = r$) loss function increase in A.  Again,  $p + q > 0$.  We summarize this as Property 1:\newline
\textbf{Property 1:}  The loss function defined by equations (1a) and (1b) increases in $A$ for any given absolute relative error.  This is assured whenever $q > -p$, or, equivalently, $p + q > 0$.

\section{Estimating the Loss Function\label{LFest}}
This Section assumes the existence of an impartial decision-maker with well-formed preferences, such an investor motivated by profits affected by prediction error.  The decision-maker is queried about his preferences to ultimately estimate the parameters of his loss function.  The steps in this include ascertaining reasonable bounds on $A$ 
and $\epsilon$, given $A$; presenting pairs $(\epsilon_j, A_j)$ to the decision-maker for evaluation; estimating $p$ and $q$; checking $p$ and $q$ for reasonableness; and testing the specification.  This procedure is in keeping with Lindley’s (1953, p. 43) interpretation of Wald (1950):
\begin{quotation}
[T]he statistician’s answer to a problem cannot be cannot be given except in close co-operation with the questioner, who is in a position to make the decisions and share their consequences with the statistician.
\end{quotation}
In this case, the problem is to compare the accuracy of different sets of predictions, as determined by the impartial decision-maker’s preferences.

	Some alternative methods are given in Subsubsection 3.2.1, but are not recommended for use by a single decision-maker.

\subsection{Reasonable Actual Values and Error Bounds}

This analysis requires that a reasonable range for $A$ and error bounds be obtained from a knowledgeable source (often, the decision-maker) beforehand.  This ensures that the analysis operates only on values which can reasonably be expected to appear.  The error bounds need to be obtained to prevent situations in which a pair $(\epsilon_j, A_j)$ is completely unacceptable to the decision-maker.  This can create difficulties as noted in the next Subsection.

\subsection{Determining the Loss Function at a Set of Points and Estimating it Globally}

The decision-maker is given a set of pairs $(\epsilon_j, A_j)$ and asked to express his satisfaction r $U(\epsilon_j, A_j)$ for each pair $(\epsilon_j, A_j)$.\footnote{The use of $U$ to denote this function corresponds to the economic concept of utility, as discussed in footnote \ref{utility} above.  The decision-maker is asked to reveal his utility, which is negated to produce loss.}  The framing of “satisfaction” is important: a possible question is “What percentage acceptability does this pair have, where 0\% is completely unacceptable and 100\% is completely accurate?”\footnote{100\% satisfaction is equivalent to $L = 0$.}  The exact phrasing of this question is a matter of survey design, which is beyond the scope of the present paper.  An answer of 0\% indicates that the pair has an unreasonable error bound.  Any pair with 0\% acceptability has to be removed from the analysis and replaced, if desired, by another pair with reasonable values.  The selection of the pairs $(\epsilon_j, A_j)$ can be done by many means.  Perhaps the simplest method is to choose a set of values of $A$, spanning its range, and then a set of errors for each value of $A$, which are well within their reasonable ranges.  The number of pairs has to be determined as well.  A small number can make the regression procedure below unreliable, while a large number can overwhelm the decision-maker.  Again, these points are all matters of survey design.

Once the $U(\epsilon_j, A_j)$ are obtained, the associated values of the loss function  $L(\epsilon_j, A_j)$ can be determined by the equation 
\begin{equation}
	L(\epsilon_j, A_j) = 100 - U(\epsilon_j, A_j)
\end{equation}
where $L$ and $U$ are expressed in terms of percentage points. We can see why points with $U = 0$ are removed: At these points $U = 100$ for some $\epsilon_j > 0$.  Higher values of $\epsilon_j$ produce the same value of $L$, in contradiction to Assumption 2.  Then equation (1b) is estimated by the regression equation formed by taking its logarithm and evaluating it at each pair $(\epsilon_j, A_j)$:\footnote{Throughout this paper, we understand logarithms and the exponent function to be natural.}
\begin{equation}
	\log L_j = p \log \epsilon_j + q \log A_j + u_j
\end{equation}
where $L_j > 0,a$ is the constant of integration, an ignorable scaling factor, and $u_j$ is the regression residual.  The values of $p$ and $q$ found from this regression become the parameters of the loss function.  One caveat is in order: if the decision-maker reports a $U_j = 100$, then $L_j = 0$ and  $\log L_j = -\infty$.  The fix for this is to either delete this observation from the analysis or to choose a small $\delta > 0$ to bound the $L_j$ from below.

\subsubsection{Multiple Decision Makers}

Within an organization, one can expect individuals’ preferences to become homogeneous over time.\footnote{Coleman (1964) models processes by which preferences within groups become more homogeneous.  Two effects (Coleman, 1964, pp. 348-349) can account for this.  “Contagion” simply involves individuals’ influencing each other, while “heterogeneity” implies self-selection among individuals comprising a group.}  Therefore, it is likely that their values of $p$ and $q$ estimated from regression equation (5) will be very similar.  However, individuals from different organizations may have very different preferences.  For example, one individual may be more willing to accept inaccuracy in small areas than another.  It is thus possible that their loss functions will be very dissimilar.  Aggregating these dissimilar preferences into a single metric becomes an exercise in group decision making.  Zahedi (2000) discusses group decision making methods. It is also quite possible that these dissimilar preferences have no substantial effects on the analysis. 

	One alternative to Zahedi’s (2000) procedures lies in constructing several loss functions.\footnote{I’d like to thank Gregg Diffendal for pointing this out.}  Each loss function can be made to obey Assumptions 1-3 and Property 1.  Since estimation is not used, the loss functions are effectively chosen arbitrarily.\footnote{This is a standard technique for evaluating census adjustments.} The ensemble of loss functions can be used to compare different sets of predictions.  If all of the loss functions are in agreement about the rankings of the prediction sets’ total losses, then one can argue that the exact parametrization does not matter: each loss function produces the same result.  However, if the loss functions disagree, then one is forced back onto the horns of subjectivity, this time in the form of a multiattribute utility decision problem.\footnote{See Sarin (2000) for a discussion of multiattribute utility.}  These require specifying weights to be assigned to each loss function or using some sort of tie-breaking function.  The essence of this problem is stated by Lindley (1985, p. 180): 
\begin{quotation}
[N]o coherent approach exists.\footnote{Lindley (1985, p. 59) defines coherence in terms of probabilities of events and utilities of their consequences.}
\end{quotation}
  The fundamental problem is that there is no way to specify a unique utility function with which to evaluate prediction errors.  Arrow’s (1950) famous Impossibility Theorem disproving the existence of a function that aggregates individuals’ preferences and obeys some weak conditions underlies the last claim.

	A final alternative is to adopt a loss function which works “well” in most cases as a convention in the manner of Keyfitz (1979).  The Mean Absolute Percentage Error (MAPE,) discussed in Section 6, is frequently used in this sense.  Another alternative is the Webster-Saint Lagüe Rule, discussed in Section 5.  The Webster-Saint Lagüe Rule is a particular parametrization of our loss functions.  That the chosen loss function need not work well in every case has been explicated by Lerner (1957, p. 441) in the context of legal rules:
\begin{quotation}
[W]ith all general rules, there are particular cases where...it would be better for the rule not to be applied.
\end{quotation}
The poor performance of a particular loss function in a particular circumstance does not invalidate its use in general.	

\subsection{Specification Tests}

While equations (1a) and (1b) are the simplest form that thid loss function can take, they are hardly the only forms.  In fact, an infinite number of forms of loss functions are possible.  One quick specification test is to look at the sum $p + q$.  If it is nonpositive, then it violates Property 1.  Either the equation is misspecified,\footnote{Remember that equations (1a) and (1b) represent only one of an infinite number of loss functions that satisfy Assumptions 1-3 and Property 1.} or the decision-maker’s preferences violate Property 1.\footnote{In some contexts, this may be reasonable.}  The form of the function may be misspecified, for which many tests are available.  Heteroscedasticity may be present, which can be accounted for by weighted regression.

\subsection{Optimal Averages of Predictions \label{optimal}}

Given two or more sets of predictions, it may be possible to improve on their accuracy by finding an optimal weighted average of these predictions.  Suppose there are two sets of estimates $\mathbf{P}^1 $ and $\mathbf{P}^2 $, with associated total losses $\mathcal{L}^1$  and $\mathcal{L}^2$.  Let $w_1$ and $w_2$ be weights such that $w_1 + w_2 = 1$. By varying the values of $w_1$   and $w_2$, one creates new sets of predictions, one per set of weights.\footnote{The new estimates may have to be constrained to sum up to a predetermined overall total or several predetermined subtotals.}  The total loss minimizing set of estimates and weights is then chosen.  The optimal search method is unclear: this probably requires grid-searching.  With $k > 2$ estimate sets, this becomes a possibly expensive mesh search in $k - 1$ dimensions.

	This can be seen to be a sort of Bayesian enterprise.  The usual practice is to chose $w_1 = w_2 = 1/2$.  This accords with the Bayesian principle of indifference.  By letting the weights vary and choosing the optimal set with regard to a given loss function, one effectively creates a Bayesian prior which places those weights on the different sets of predictions.  The practice of putting weights on sets of predictions is a form of model averaging.\footnote{For a fuller explanation of Bayesian model averaging see Hoeting, Madigan and Volinsky (1999).  Many machine learning algorithms are model averaging algorithms in that they create ensembles of estimators, then average over all ensemble members to produce estimates or predictions.}  The weight put on a particular set of predictions is often used as the weight for the model which produced those predictions. The averaged model is then used to generate new sets of predictions.  Suppose there are two sets of estimates $\mathbf{P}^1 $ and $\mathbf{P}^2 $, with associated total losses $\mathcal{L}^1$  and $\mathcal{L}^2$.  Let $w_1$   and $w_2$ be weights such that $w_1$   and $w_2 = 1$.\footnote{This argument generalizes to any number $k$ sets of estimates with $\sum_{i=1}^k w_i = 1$.} By varying the values of $w_1$   and $w_2$, one creates new sets of predictions of the form $\lbrace P^{1,2}_i = w_1P^1_i + w_2P^2_i\rbrace$  and computes the total loss associated with each set of weights.\footnote{The new estimates may have to be constrained to sum up to a predetermined overall total or several predetermined subtotals.  See Subsection \ref{apriori}.} The total loss minimizing set of estimates and weights is then chosen.  The optimal search method is unclear: this probably requires grid-searching.  With $k >2$ prediction sets, this becomes a possibly expensive mesh search in $k - 1$ dimensions.

\section{Predictions as Apportionments: The Webster-Saint Lag\"ue Rule \label{wsl}}

Often predictions represent apportionments, that is, resource or legislative seat allocations.  The U.S. decennial census is used to apportion the House of Representatives.  Population estimates are used by some states to apportion tax revenues to localities.  One would like to have the rule used for evaluating the predictions satisfy various fairness criteria.  Balinski and Young (2001) show that the Webster-Saint Lag\"ue Rule satisfies a large number of fairness criteria.\footnote{However, while Arrow’s(1950) Impossibility Theorem implies that no voting system can simultaneously satisfy all fairness criteria, the Webster-Saint Lag\"ue Rule is the only apportionment rule considered by Balinski and Young (1979, 2001) and Ernst (1992) that can be represented by minimizing the sum of loss functions satisifying Assumptions 1-3 and Property 1 (Spencer, 1985).  I would like to thank Gregg Diffendal for informing me of Ernst (1992).}  Most of these criteria result from the Webster-Saint Lag\"ue Rule being a divisor method.  The Webster-Saint Lag\"ue Rule  is the only divisor method considered  by Balikski and Young which has an admissible (in the present paper) loss function representation:  $p = 2$ and $q = -1$, (Balinski and Young, 2001) See Subsection \ref{apportion} below for other apportionment-based loss functions.  An additional normative criterion of particular interest is unbiasedness: for any two disjoint groups of areas, with all members in one group larger than those in the other, the probability that the method favors large areas equals the probability that it favors the small areas. (Balinksi and Young, 2001, Theorem 5.3)  

	Since an optimal apportionment minimizes $\mathcal{L}_W$, it is natural to use it as a measure of accuracy.  The greater its value, the greater the departures from the normative criteria used to evaluate the predictions.  This way of characterizing loss avoids the need for an impartial decision-maker.  It also has an a priori probabilistic interpretation, as shown in Subsection {\ref{apriori}}.

Fellegi (2008) effectively proposed a variant of the Webster-Saint Lag\"ue Rule.  This is studied in Subsection \ref{fellegi}.

\subsection{An A Priori Probabilistic Interpretation of the Webster-Saint Lag\"ue Rule \label{apriori}}

Let the sets of predictions be indexed by $j$.  Let the expected mean squared error of the prediction from set $j$ for area $i$ be proportionate to $A_i$: $\mathbf{E}(P_i^j-A_i)^2 = cA_i$.  If the predictions are unbiased, $\mathbf{E}(P_i^j-A_i)=0$, Webster’s rule actually estimates the underlying variances,  thereby enabling a ranking based on the average estimated variance.  This is proved below.

	Let the estimates be unbiased (i.e., $\mathbf{E}\epsilon_i^j = 0$ for all $i$ and $j$) and let $P_i^j$ be the sum of $A_i$ independent variables $\eta_i^j$ with common mean $\mu$ and variance $\sigma^2_j$.  Then $\mathbf{E}P_i^j=A_i$  and 
\begin{subequations}
\begin{align}
	\mathbf{E}(P_i^j-A_i)^2 &=  \mathbf{E}(A_i + \epsilon_i^j - A_i)^2 \\
	&= \mathbf{E}(\epsilon_i^j )^2 \\
	&= \mathbf{E}\sum_{k=1}^{A_i} \eta_i^j \\
	&=  \mathrm{Var} \left(\sum_{k=1}^{A_i} \eta_i^j\right)  \label{nobias} \\
	&= A_i  \mathrm{Var}(\eta_i^j)  \label {vareq}\\
	&= A_i \sigma^2_j \label{finalsum},
\end{align}
\end{subequations}
where equation (\ref{nobias}) follows from asssumed unbiasedness, equation (\ref{vareq}) follows from the variance of the sum of independent random variables being equal to the sum of their variances (equal in this example) and equation (\ref{finalsum}) follows from the definition of $\sigma^2_j$.  Therefore,
\begin{subequations}
\begin{align}
	\mathbf{E}\frac{(P_i^j-A_i)^2}{A_i} &= \frac{A_i \sigma^2_j}{A_i} \\
	&= \sigma^2_j .
\end{align}
\end{subequations}
Thus, the  Webster-Saint Lag\"ue rule produces estimates of the variances of each observation, $A_i \hat{\sigma}_{ij}^2$.  Using its minimization to select predictions chooses the estimates set with the lowest $\hat{\sigma}_j^2$.  Relaxing the the identical variance assumption so that $\sigma^2_{ij}$ is not constant causes this rule to estimate the average variance $\hat{\bar{\sigma}}^2_j = \frac{1}{n}\sum_{k=1}^{n} \hat{\sigma}_{kj}^2$. Now, the Webster-Saint Lag\"ue loss minimization selects the predictions set with the lowest average estimated variance.  This is robust to small departures from the assumptions. 

\subsection{Fellegi's Variant\label{fellegi}}

Fellegi (1980, eq. 8, 194) proposes an inequality measure based on differences between optimal and predicted shares.  He uses conventions to assign exponents.  Hogan and  Mulry (2014, eq. 94, p. 124) show that Fellegi's measure is equivalent to the loss function\footnote{This is after removing constants and the summation.}
\begin{equation}
	\frac{1}{\sum_I A_i}\left(\frac{P_i}{\sum_i P_i} - \frac{A_i}{\sum_i A_i}\right)^2.
\end{equation}
It is easy to show that this obeys Assumption 3, $\partial\ell/\partial A > 0$  for all $A > 0$, is satisfied.  However, Assumptions 1 and 2 have to be modified.
Letting
\begin{equation}
	\epsilon'_i = \left|\frac{P_i}{\sum_i P_i} - \frac{A_i}{\sum_i A_i}\right|
\end{equation}
and substituting into Assumptions 1 and 2 yields new assumptions.  Supressing subscripts, these become:
\textbf{Assumption $\mathbf{1'}$:} $L(A + \epsilon'; A) = L(A - \epsilon'; A)$ for all $A > 0$.\newline
and \newline
\textbf{Assumption $\mathbf{2'}$:}   $\partial\ell/\partial\epsilon' > 0$ for all $\epsilon' > 0$.

Fellegi's loss function satisfies the two new assumptions.  Noting that $\ell(\epsilon',A) = (\epsilon')^2 A^{-1}$, we see that $p = 2$ and $q=-1$, thereby satisfying Property 1.  The key difference between the Fellegi and Webster-Saint Lag\"ue loss functions is that the former defines errors as differences in shares while the latter uses differences in levels.

\section{Example of Evaluating Predictions Using Loss Functions\label{example}}

Loss functions can produce entirely different and more meaningful results than common error measures such as the mean absolute percentage error (MAPE).\footnote{That is the loss function is computed with $p = 1$ and $q = -1$, absolute values taken, then multiplied by 100 for reexpression in terms of percentage points.} Table 1 shows the true values of six areas, $A_i, i = 1,\dots,6$, and three sets (Scenarios) of absolute errors $(\epsilon_i)$, along with the corresponding absolute percentage errors ($\mathrm{APE}_i)$ and Webster-Saint Lag\"ue Rule loss function values $L_i$.\footnote{This example is from Coleman (2000).}  The bottom row shows the means of the last two variables.  These are simply MAPE and $\mathcal{L}_W/n$, respectively.

\begin{table}[]
\caption{Comparison of MAPE and Webster-Saint Lagüe Loss Functions for Three Hypothetical Scenarios}
\vspace*{\bigskipamount}
\begin{tabular}{lr|rrr|rrr|rrr|}
\cline{3-11}
                            & \multicolumn{1}{l|}{} & \multicolumn{3}{c|}{Scenario 1}                                            & \multicolumn{3}{c|}{Scenario 2}                                    & \multicolumn{3}{c|}{Scenario 3}                                      \\ \hline
\multicolumn{1}{|l|}{Area}  & $A_i$                    & \multicolumn{1}{r|}{$\epsilon_i$} & \multicolumn{1}{r|}{$\mathrm{APE}_i$} & $L_i$ & \multicolumn{1}{r|}{$\epsilon_i$} & \multicolumn{1}{r|}{$\mathrm{APE}_i$} & $L_i$  & \multicolumn{1}{r|}{$\epsilon_i$} & \multicolumn{1}{r|}{$\mathrm{APE}_i$} & $L_i$    \\ \hline 
\multicolumn{1}{|l|}{1}     & 100000                & \multicolumn{1}{r|}{2000}        & \multicolumn{1}{r|}{2}    & 40.00         & \multicolumn{1}{r|}{1000}        & \multicolumn{1}{r|}{1}    & 10.0  & \multicolumn{1}{r|}{3000}        & \multicolumn{1}{r|}{3.0}    & 90.00    \\ \hline
\multicolumn{1}{|l|}{2}     & 50000                 & \multicolumn{1}{r|}{1000}        & \multicolumn{1}{r|}{2}    & 20.00          & \multicolumn{1}{r|}{500}         & \multicolumn{1}{r|}{1}    & 5.0   & \multicolumn{1}{r|}{850}         & \multicolumn{1}{r|}{1.7}  & 14.45 \\ \hline
\multicolumn{1}{|l|}{3}     & 10000                 & \multicolumn{1}{r|}{200}         & \multicolumn{1}{r|}{2}    & 4.00           & \multicolumn{1}{r|}{100}         & \multicolumn{1}{r|}{1}    & 1.0   & \multicolumn{1}{r|}{170}         & \multicolumn{1}{r|}{1.7}  & 2.89  \\ \hline
\multicolumn{1}{|l|}{4}     & 5000                  & \multicolumn{1}{r|}{100}         & \multicolumn{1}{r|}{2}    & 2.00           & \multicolumn{1}{r|}{50}          & \multicolumn{1}{r|}{1}    & 0.5 & \multicolumn{1}{r|}{85}          & \multicolumn{1}{r|}{1.7}  & 1.45  \\ \hline
\multicolumn{1}{|l|}{5}     & 1000                  & \multicolumn{1}{r|}{20}          & \multicolumn{1}{r|}{2}    & 0.40         & \multicolumn{1}{r|}{10}          & \multicolumn{1}{r|}{1}    & 0.1 & \multicolumn{1}{r|}{17}          & \multicolumn{1}{r|}{1.7}  & 0.29  \\ \hline
\multicolumn{1}{|l|}{6}     & 100                   & \multicolumn{1}{r|}{2}           & \multicolumn{1}{r|}{2}    & 0.04        & \multicolumn{1}{r|}{10}          & \multicolumn{1}{r|}{10}   & 1.0   & \multicolumn{1}{r|}{2}           & \multicolumn{1}{r|}{2.0}    & 0.04  \\ \hline
\multicolumn{1}{|l|}{Means} &                       & \multicolumn{1}{r|}{}            & \multicolumn{1}{r|}{2}    & 11.08       & \multicolumn{1}{r|}{}            & \multicolumn{1}{r|}{2.5}  & 2.9 & \multicolumn{1}{r|}{}            & \multicolumn{1}{r|}{1.97} & 18.19 \\ \hline
\end{tabular}
\end{table}

	Three Scenarios are used to compare the results of an evaluation using a loss function to those obtained by using MAPE.  Scenario 1 is the baseline scenario with $\mathrm{APE}_i) \equiv 2$ and $\mathcal{L}_W/n = 11.08$.  In Scenario 2, $\mathrm{APE}_i$ is reduced to 1 for $i \le 5$, but $\mathrm{APE}_6$ increases to 10.  That is, all but the smallest areas have their APEs halved, but the very smallest area’s APE increases by a factor of 5.  MAPE increases to 2.5, but $\mathcal{L}_W/n$ falls to 2.9. Thus, MAPE ranks Scenario 2 as being less accurate than Scenario 1, even though the individual errors are smaller except for the very smallest area. On the other hand, using a loss function takes into account the size of the smallest area and discounts its accuracy loss and considers Scenario 2 to be more accurate. In Scenario 3, $\mathrm{APE}_i$ falls by 15\% to 1.7 for $i = 2  \le i \le 5$, rises by 50\% to 3 in the largest area, and is unchanged in the smallest area.  MAPE falls slightly to 1.97, but $\mathcal{L}_W/n$ rises to 18.19. Thus, MAPE considers Scenario 3 to be superior to Scenario 1, as a result of the general reduction in the $\mathrm{APE}_i$, in spite of the major loss in accuracy in the largest area. The  Webster-Saint Lag\"ue Rule loss function, on the other hand, puts a large weight on the accuracy loss in area 1 and increases its error measure relative to Scenario 1.  Thus, this loss function puts increasing weight on an error as the size of the area increases.  Putting all of these together, we find that MAPE and the Webster-Saint Lag\"ue loss function produce exactly opposite rankings of the Scenarios.

	The reader may notice that the losses have no intuitive interpretation, unlike the absolute percentage errors that go into MAPE.  This is hardly unknown to this enterprise.  For example, Lindley (1953, p. 46) writes
\begin{quote}
	...they have no direct interpretation in the real world...
\end{quote}
This quote refers to “weight” functions, Wald’s (1950) original term for loss functions.\footnote{Wald (1950, p. 8) introduces weight functions in a subsection entitled “Losses Due to Possible Wrong Terminal Decisions and Cost of Experimentation.”}  The upshot of this statement is that one should generally not expect loss functions to coincide with any easily interpreted measures.

\section{Comparison to Other Metrics \label{compare}}

This Section compares other, commonly used, metrics to the metric defined by loss functions (1a) and (1b) and Property 1.

\subsection{Root Mean Squared Error}

The root mean squared error (RMSE) is defined as $\sqrt{\frac{\sum_{i=1}^n (P_i-A_i)^2}{n}}$. This is equivalent to $p = 2$ and $q = 0$.  This violates Assumption 3.  It is similar to the standard deviation and is only useful if the expected mean squared errors, $\mathbf{E}\epsilon_i^2$, are constant.  This clearly is not true of most cross-sectional predictions, which often span several orders of magnitude.

\subsection{Median Absolute Percentage Error}

The median absolute percentage error (MedAPE) is a close relative of MAPE and, for our purposes, can only be defined by its total loss function, $100 *\mathrm{median}_i \lvert \frac{P_i -A_i}{A_i} \rvert$.  Not only does it have the disadvantages of MAPE, but its use of the median instead of the mean makes it insensitive to large errors.\footnote{It should also be noted that additive separability and the von Neumann-Morgenstern expected utility axioms are violated as well.  See footnote \ref{utility}.  Coleman (2025) further clarifies this point by showing that quantile total loss functions (e.g., MedAPE and 90PE of the next Subsection) in general have unbounded insensitivity to outliers.}

\subsection{90\textsuperscript{th} Percentile Absolute Error}

Smith and Sincich (1992) use the 90\textsuperscript{th} Percentile Absolute Percentage Error (90PE) as one of their metrics for comparing cross-sectional forecasts.  Like MedAPE, this can only be defined in our framework by the total loss function, $100 * \lvert \frac{P_{90} -A_{90}}{A_{90}} \rvert$, where the subscript 90 refers to the observation whose absolute percentage error is greater than that of 90\% of all the observations.  90PE shares MedAPE’s weaknesses, while being sensitive to fewer extreme errors.

\subsection{Root Mean Squared Percentage Error}

The Root Mean Squared Percentage Error (RMSPE) is defined as $\sqrt{\frac{[\sum_{i=1}^n (P_i-A_i)/{\sum_{i=1}^n A_i}]^2}{n}}$.  This is equivalent to $p = 2$ and $q = -2$.  Like MAPE, this violates Property 1. It is useful only when the a priori relative mean squared errors are constant.  That is, $\mathbf{E}(\frac{P_i - A_i}{A_i})^2=c$, for some constant $c > 0$.

\subsection{Apportionment-Based Loss Functions\label{apportion}}

Spencer (1985) develops five apportionment-based loss functions.  In addition to the Webster-Saint Lagüe Rule, these are (1) $p = 1, q = 0$, or more generally, $p > 0, q = 0$ loss function. (Hamilton’s method), (2) $p = 2, q = -1$ with $P_i$ raised to the $q$ instead of $A_i$ (Hill’s method/Method of Equal Propertions), (3)  $\max_i P_i/A_i$ (Jefferson/d'Hondt method) and (4) $\max_i A_i/P_i$ (Adam’s Method).  Hamilton’s and Hill’s methods both violate Assumption 3.  The Jefferson/d'Hondt method violates both Assumption 3 and the von Neumann-Morgenstern axioms.  Adam’s Method additionally violates Assumptions 1 and 2.  Spencer (1986, p. 398) advocates incorporating expectations into these methods to measure accuracy.  However, since we have assumed the distributions generating the errors to be unknown, this is impossible.

\subsection{Some Examples which Obey Assumptions 1-3 and Property 1}

Bryan (2000) proposes using $p = 1$ and $q = \log(range)/25 - 1$ in the context of small-area population estimates evaluation, where $range$ is the range of the $A_i$.  This obeys Assumption 3 and Property 1 for $range < \exp(25) \approx  72{,}000{,}000{,}000$.  Apparently, $p = 1$ and $q = -1/2$ was proposed for evaluating the U.S. Census Bureau’s Small Area Income and Poverty Estimates (William R. Bell, personal correspondence).

\section{Loss Function-Based Bias Measures \label{bias}}

Any of our loss functions can be converted to a bias measure simply by multiplying loss by the sign of $\epsilon$, similar to the construction of the Mean Algebraic Percentage Error (MALPE) from MAPE. From page \pageref{epsilon} we know that $\epsilon = |P - A|$. Therefore, equation (\ref{PAloss}) becomes
	\begin{equation}
		S_i(P,A) = \sgn (\epsilon_i) |P-A|^p A^q,
           \end{equation}
where $S_i$ is interpreted as the signed loss function.  Besides being a measure of bias, it can rank observations by how overprediction or underprediction can contributes to the loss of accuracy.  Using it with external data may enable identification of factors affecting accuracy.\footnote{Judson, Popoff and Batutis (1999) do this for the U.S. Census Bureau's 1990 county population estimates using the Algebraic Percentage Error (ALPE) for their bias measure.  ALPE is defined as $100 * S_i(P,A)$ with $p = 1$ and $q  = -1$.  Taking its mean also results in MALPE.}

\section{Conclusion \label{conclude}}

This paper has axiomatically developed loss functions for measuring the accuracy of cross-sectional predictions.  It is a generalization of several metrics in the literature and avoids the pitfalls common to most of them when applied to cross-sectional data.  When the predictions represent resource allocation levels, the Webster-Saint Lagüe Rule provides an exact parametrization. Other cases require a great deal of information from an impartial decision-maker and are not immune to manipulation.  A forecaster can easily manipulate their parameters to make his forecasts look better than those of their competitors.  Decision-makers with interests concentrated in particular size ranges can degrade performance with respect to other size ranges.  Therefore, this approach should only be used by disinterested parties who wish to evaluate the overall performance of cross-sectional predictions. Finally, a bias measure has been derived that can be used to measure bias or possibly identify factors affecting accuracy when combined with other data.

\section*{References}

Armstrong, J.S. (1985).  Long-Range Forecasting:  From Crystal Ball to Computer, 2nd edition.  New York:  Wiley.
\newline\newline
Armstrong, J.S. and Collopy, F. (1992). Error Measures for Generalizing about Forecasting Methods: Empirical Comparisons. International Journal of Forecasting, 8, 69-80.
\newline\newline
Arrow, K.J. (1950). A Difficulty in the Concept of Social Welfare. The Journal of Political Economy, 58, 328-346.
\newline\newline
Balinski, M.L. and Young, H.P. (1979).  Criteria for Proportional Representation.  Operations Research, 27, 80-95.
\newline\newline
Balinski, M.L. and Young, H.P. (2001). Fair Representation. Second Edition.  Washington, D.C.: Brookings Institution Press.
\newline\newline
Beaumont, P.M. and Isserman, A.M. (1987).  Tests of Accuracy and Bias for County Population Projections: Comment. Journal of the American Statistical Association, 82, 1004-1009.
\newline\newline
Bryan, T. (1999).  U.S. Census Bureau Population Estimates and Evaluation with Loss Functions. Statistics in Transition, 4, 537-548.
\newline\newline
Coleman, C.D. (2000).  Evaluation and Optimization of Population Projections Using Loss Functions.  In Federal Forecasters Conference 2000: Papers and Proceedings, ed. Gerald, D.E., Washington: Department of Education, Office of Educational Research and Improvement. \url{https://cer.columbian.gwu.edu/sites/g/files/zaxdzs4246/files/downloads/FFC2000_reduced.pdf}
\newline\newline
Coleman, C D. (2025).  Total Loss Functions for Assessing the Accuracy of Cross-Sectional Estimates and Forecasts. \url{https://doi.org/10.48550/arXiv.2507.15136}
\newline\newline
Coleman, J.S. (1964).  Introduction to Mathematical Sociology. New York: Free Press of Glencoe.
\newline\newline
Davis, Sam T. (1994). Evaluation of Postcensal County Estimates for the 1980s. Population Division Working Paper No. 5, Washington, D.C.: U.S. Census Bureau.  \url{http://www.census.gov/population/www/documentation/twps0005/twps0005.html}
\newline\newline
Ernst, L.R. (1992). Apportionment Methods for the House of Representatives and the Court Challenges.  SRD Research Report 92/06, Washington, D.C.: U.S. Census Bureau.  \url{http://www.census.gov/srd/papers/pdf/rr92-6.pdf}
\newline\newline
Fellegi, I. (1980). Should the Census Count Be Adjusted for Allocation Purposes—Equity Considerations. Conference on Census Undercount: Proceedings of the 1980 Conference. U.S. Department of Commerce, Washington, DC,  193-203.
\newline\newline
Fildes, R. (1992).  The Evaluation of Extrapolative Forecasting Methods.  International Journal of Forecasting, 8, 81-98.
\newline\newline
Hastie, T., Tibshirani, R. and Friedman, J. (2001). The Elements of Statistical Learning: Data Mining, Inference, and Prediction: With 200 Full-Color Illustrations. New York : Springer.
\newline\newline
Hoeting, J.A., Madigan, D., Raftery, A.E. and Volinsky C.T. (1999). Bayesian Model Averaging: A Tutorial (with discussion.  Statistical Science, 14, 382-417. \url{https://projecteuclid.org/journals/statistical-science/volume-14/issue-4/Bayesian-model-averaging--a-tutorial-with-comments-by-M/10.1214/ss/1009212519.full}
\newline\newline
Hogan, H. and Mulry, M. H. (2014). “Assessing Accuracy of Postcensal Estimates: Statistical Properties of Different Measures,” in N. Hogue (Ed.), Emerging Techniques in Applied Demography. Springer. New York.
\newline\newline
Judson, D., Popoff, C. and Batutis, M. (1999). "An Evaluation of the Accuracy of U.S. Bureau of the Census County Population Estimates."  \url{https://www.census.gov/library/working-papers/1999/demo/judson-01.html}
\newline\newline
Keyfitz, N. (1979). Information and Allocation: Two Uses of the 1980 Census. The American Statistician, 33, 45-50. 
\newline\newline
Lerner, A.P. (1957).  The Backward-Leaning Approach to Controls.  The Journal of Political Economy, 65, 437-441.
\newline\newline
Lindley, D.V. (1953). Statistical Inference. Journal of the Royal Statistical Society, Series B, 15, 30-76.
\newline\newline
Lindley, D.V. (1985). Making Decisions, 2nd ed.,  New York: Wiley.
\newline\newline
Muhsam, H.V. (1966). The Use of Cost Functions in Making Assumptions for Population Forecasts. In United Nations, World Population Conference, 1965, Volume III. New York: United Nations, 23-26.
\newline\newline
Mulry, M.H. and Spencer, B.D. (1993).  Accuracy of the 1990 Census and Undercount Adjustments.  Journal of the American Statistical Association, 88, 1080-1091.
\newline\newline
National Research Council (1980). Estimating Population and Income of Small Areas. Washington, D.C.: National Academy Press.
\newline\newline
Sarin, R.K. (2000).  Multi-Attribute Utility Theory.  In Encyclopedia of Operations Research and Management Science, 2nd ed., eds. Gass, S.I. and Harris, C.M., Boston: Kluwer Academic Publishers, 526-529.
\newline\newline
Smith, S.K. (1987). Tests of Accuracy and Bias for County Population Projections. Journal of the American Statistical Association, 82, 991-1003.
\newline\newline
Smith, S.K. and Sincich, T. (1992). Evaluating the Forecast Accuracy and Bias of Alternative Population Projections for States. International Journal of Forecasting, 8, 495-508.
\newline\newline
Spencer, B.D. (1980).  Benefit-Cost Analysis of Data Used to Allocated Funds.  New York: Springer-Verlag.
\newline\newline
Spencer, B.D. (1985). Statistical Aspects of Equitable Apportionment. Journal of the American Statistical Association, 80, 815­­­-822.
\newline\newline
Spencer, B.D. (1986). Conceptual Issues in Measuring Improvement in Population Estimates.  In Bureau of the Census, Second Annual Research Conference: Proceedings, March 23-26, 1986, 393-407.
\newline\newline
Stanford Research Institute (1974).  General Revenue Sharing Data Study, Volume III.  Menlo Park, CA:  Stanford Research Institute.
\newline\newline
Tayman, J., Schafer E., and Carter L. (1998). The Role of Population Size in the Determination and Prediction of Population Forecast Errors: An Evaluation using Confidence Intervals for Subcounty Areas. Population Research and Policy Review, 17, 1-20.
\newline\newline
Wald, A. (1950). Statistical Decision Functions, New York: Wiley.
\newline\newline
Zahedi, F. (2000). Group Decision Making.  In Encyclopedia of Operations Research and Management Science, 2nd ed., eds. Gass, S.I. and Harris, C.M., Boston: Kluwer Academic Publishers, 343-350.

\end{document}